\documentclass[%
 reprint,
 amsmath,amssymb,
 aps,
]{revtex4-2}

\usepackage{graphicx}% Include figure files
\usepackage{dcolumn}% Align table columns on decimal point
\usepackage{bm}% bold math
\usepackage{siunitx}
\usepackage{comment}

\usepackage{hyperref}% add hypertext capabilities
\usepackage[mathlines]{lineno}% Enable numbering of text and display math
\usepackage{xcolor}
\newcommand\lr[1]{\left\langle #1 \right\rangle}

\usepackage[utf8]{inputenc} % allow utf-8 input
\usepackage[T1]{fontenc}    % use 8-bit T1 fonts
\usepackage{hyperref}       % hyperlinks
\usepackage{cleveref}
\usepackage{url}            % simple URL typesetting
\usepackage{booktabs}       % professional-quality tables
\usepackage{amsfonts}       % blackboard math symbols
\usepackage{nicefrac}       % compact symbols for 1/2, etc.
\usepackage{microtype}      % microtypography
\usepackage{lipsum}

\usepackage{algorithm2e}

\SetKwInOut{Parameter}{Parameters}

\usepackage{subcaption}

\def\R{\mathbb{R}}
\DeclareSIUnit\pN{\pico\newton}
\DeclareSIUnit{\us}{\micro\second}
\DeclareSIUnit[per-mode=symbol]\D{\um\squared\per\second}
\DeclareSIUnit[inter-unit-product=\ensuremath{{}\cdot{}}]\pNnm{\pico\newton\nano\meter}

\newcommand*{\kT}{\, k_{\textup{B}}\textup{T}}

\newcommand{\bias}{\tilde{B}\left(\hat{\mathbf{F}}\right)}

\begin{document}

 \title{Iterative improvement of free energy landscape reconstructions with optimal protocols derived from differentiable simulations} 

 \author{Oliver Cheng$^{1,\dagger}$} \author{Zofia Adamska$^{2,\dagger}$} \author{Michael P. Brenner$^{1}$} \author{Megan C. Engel$^{1,3}$}  
 \affiliation{$^1$School of Engineering and Applied Sciences, Harvard University, Cambridge, MA 02138} \affiliation{$^2$Division of Physics, Mathematics and Astronomy, California Institute of Technology, Pasadena, CA 91125} \affiliation{$^3$Centre for Molecular Simulation, Department of Biological Sciences, University of Calgary, Calgary, Canada, T2N 1N4}
 \affiliation{$\dagger$ These authors contributed equally to this work.}
 
 %\email{olcheng712@gmail.com} 
 %\email{zadamska@perimeterinstitute.ca} 
 \email{megan.engel@ucalgary.ca} 
 %\email{brenner@seas.harvard.edu}

% \author{
% Zofia Adamska \\
% Division of Physics, 
% Mathematics and Astronomy \\
% Caltech\\
% Pasadena, CA 91125\\
%   \texttt{zadamska@caltech.edu} \\
%   \And
% Oliver Cheng \\
% School of Engineering and Applied Sciences\\
% Harvard University\\ 
% Cambridge, MA 02138 \\
% Mountain View CA 94043\\
%   \texttt{ocheng@college.harvard.edu} \\
%   \And
% Megan C. Engel \\
% School of Engineering and Applied Sciences\\
% Harvard University\\ 
% Cambridge, MA 02138 \\
%   \texttt{mcengel@seas.harvard.edu} \\
%   %% Change this to Megan's new addr at some point.
%   \And
% Michael P. Brenner \\
% School of Engineering and Applied Sciences\\
% Harvard University\\ 
% Cambridge, MA 02138 \\
% Google Research\\
% Mountain View CA 94043\\
%   \texttt{brenner@seas.harvard.edu} \\
% }

\begin{abstract} Free energy landscapes encode the kinetics, intermediates, and transition states that govern molecular processes and are thus a key target of single biomolecule research. Typical approaches to deriving optimal, error-minimizing, non-equilibrium driving protocols for estimating these landscapes require \emph{a priori} knowledge of the landscape. Here, we present an alternative: an iterative algorithm for optimizing full free energy landscape reconstructions which can be used alongside experiments on \emph{unknown} landscapes. Our approach (i) takes experimental or simulated trajectory data; (ii) reconstructs an `approximate' energy landscape; (iii) derives optimal control protocols from low-dimensional differentiable Brownian dynamics simulations on the candidate landscape using automatic differentiation; (iv) re-runs the experiment or simulation using the updated protocol; and (v) iterates until convergence. Using this approach, we recover known benchmarks from the literature and probe far-from-equilibrium regimes for symmetric, asymmetric, and triple-well energy landscapes under both 1- and 2-dimensional control. Our control protocols -- derived with no \emph{a priori} knowledge of the energy landscape -- yield substantially reduced variance and bias in free energy landscape reconstructions compared to naive linear protocols.

%[key contributions: triple well and asymmetric landscapes, iterative method for when landscape is a priori unknown, applicable in any regime, efficient because of gradient,  full energy landscape reconstructions]
\end{abstract}

\maketitle
\section{Introduction}
Calculating free energy differences is a key goal in diverse scientific contexts, from identifying and ranking drug candidates~\cite{Schindler2020DD, Bhati2022DD, klimovich_guidelines_2015, ross_maximal_2023}, to mapping the folding landscapes of disease-causing biomolecules~\cite{Harris2007FES,camilloni_energy_2012}, to elucidating enzyme catalysis mechanisms~\cite{Hayashi2010ATP}. Performing free energy calculations at equilibrium, without enhanced sampling in some form, is intractable for many systems that exhibit slow dynamics~\cite{klimovich_guidelines_2015,woodside_reconstructing_2014}. Advances in nonequilibrium statistical physics -- namely, the Crooks theorem~\cite{crooks_entropy_1999} and the Jarzynski equality~\cite{jarzynski_nonequilibrium_1997} -- provide a roadmap for computing equilibrium free energies from an ensemble of nonequilibrium experiments and simulations. While these tools have been successfully leveraged to calculate free energies in a range of contexts~\cite{truong_probing_2018,Hayashi2010ATP,nicolini_hummer_2010,Crooks2015Ising,Vanden-Eijnden2017GeometrySpin, an_experimental_2015}, a crucial limitation lies in the poor convergence behavior of the unidirectional estimator: often an impractically high number of trajectories are required to produce a reliable free energy estimate~\cite{arrar_accurate_2019, gore_bias_2003,jarzynski_rare_2006,YungerHalpernJarzynski2016}. 

There has been a recent surge of interest in approaches that improve the accuracy of Jarzynski-derived free energy calculations~\cite{LiTu2021LinearNoneq,ZhongDeWeese2024,Whitelam2025Noise,Whitelam2024PRE,blaber_skewed_2020}. A closely related problem is that of devising external protocols that minimize the dissipated work in a nonequilibrium process. This is crucial because the number of trajectories required for the unidrectional Jarzynski free energy estimate to converge is exponential in the dissipated work~\cite{gore_bias_2003,YungerHalpernJarzynski2016} as samples in the left tail of the work distribution dominate the estimator~\cite{jarzynski_rare_2006,YungerHalpernJarzynski2016}. This optimal control problem has been extensively studied and remains a topic of interest~\cite{blaber_optimal_2023,GelmanMeng1998OC}. In the near-equilibrium limit, solutions are geodesics in the space of thermodynamic geometry~\cite{GelmanMeng1998OC,sivak_thermodynamic_2016,blaber_efficient_2022,Zhong2024ThermoGeomOC}, but such general analytical solutions are intractable for all but the simplest systems ~\cite{schmiedl_optimal_2007,blaber_optimal_2023}. Outside of the linear response regime, numerical approaches are necessary to make progress~\cite{geiger_optimum_2010, ZhongWorkDissipation2022, ZhongDeWeese2024, whitelam_demon_2023,engel_optimal_2023}. For example, Whitelam \textit{et al.}~\cite{whitelam_demon_2023} used genetic algorithms to train neural-network parameterized optimal protocols for a variety of physical systems. Zhong \textit{et al.} used a trajectory reweighting scheme to compute bidirectional optimal protocols, which vastly outperform unidirectional free energy estimates~\cite{ZhongDeWeese2024}. Engel \textit{et al.} used automatic differentiation (AD) to compute minimum dissipation protocols via stochastic gradient descent~\cite{engel_optimal_2023}. Excepting the work of Whitelam \textit{et al.}~\cite{whitelam_demon_2023} and Zhong \textit{et al.}~\cite{ZhongDeWeese2024}, techniques for computing optimal control protocols on bistable landscapes require the mathematical form of the energy landscape to be known \emph{a priori}. Furthermore, existing approaches to maximizing the accuracy of nonequilibrium free energy estimates are typically tailored to calculating a \emph{single} free energy difference $\Delta F$ rather than full free energy \emph{landscapes} $F(x)$; focus on a one-dimensional control protocol (usually the position of a harmonic potential as a function of time); and treat bistable, symmetric energy landscapes. Here, we address each of these gaps.

\begin{figure*}
    \centering
    \includegraphics[width=\textwidth]{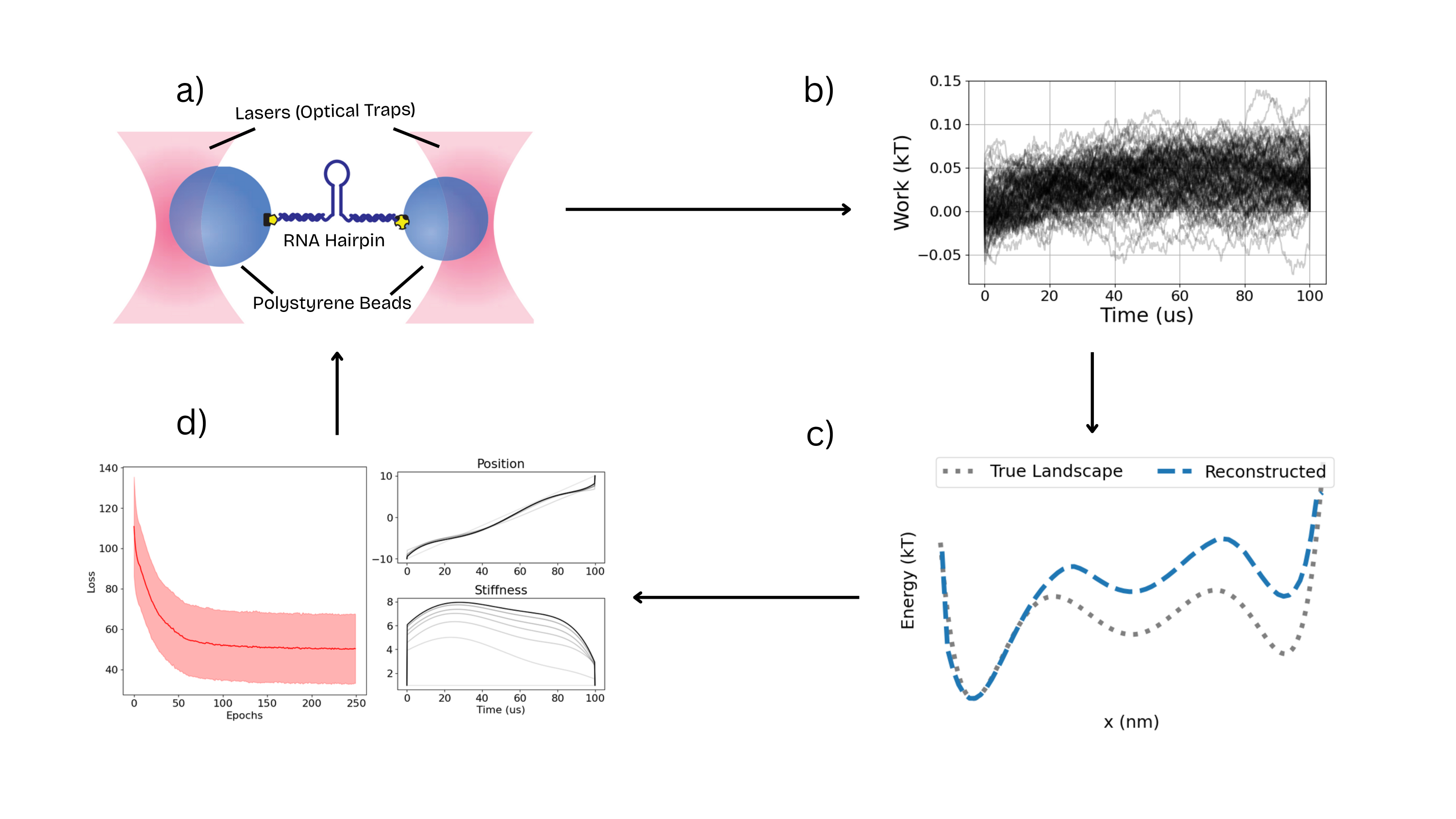}    
    \caption{Schematic of our iterative scheme for reconstructing \emph{a priori} unknown free energy landscapes. (a) Nonequilibrium barrier crossing trajectories are collected, either through simulations or experiments. Shown here is a hypothetical laser optical tweezers (LOT) setup measuring the unfolding of a DNA hairpin. Reproduced from Ref.~\cite{Woodside2016} with permission. (b) For each barrier crossing trajectory, external work is measured as a function of time. The process is stochastic, as seen by the width of the work distributions. (c) Using Equation \ref{eq:hummerszabo}, the trajectories are used to reconstruct an estimated underlying free energy landscape as a function of a molecular coordinate. (d) Brownian simulations are performed in JAX MD and automatically differentiated as described in section \ref{sec:opt_barrier_crossing} and reference~\cite{engel_optimal_2023} to minimize loss, which here is the dissipated work. This yields optimal control protocols for the time-dependent stiffness and position of the external harmonic forcing potential. Using these new control protocols, the experiments or simulations are performed again and the process repeats from (a) until the landscape estimate converges.}
    \label{fig:algorithm}
\end{figure*}

Inspired by single-biomolecule experiments that seek to reconstruct full free-energy landscapes as a function of a molecular coordinate of interest~\cite{Preiner_2007,Gebhardt_2010,GuptaWoodsideExperimental2011,engel_reconstructing_2014,woodside_reconstructing_2014}--typically, molecular end-to-end extension--we extend the approach of Engel \textit{et al.}~\cite{engel_optimal_2023} to the experimentally relevant situation wherein the molecular free energy landscape is \textit{a priori} unknown and simulations for which the \emph{gradients} of the work values cannot be easily obtained. This allows experimentalists and simulators to calibrate protocols to enhance sampling of their systems without requiring any new technology. In particular, we propose the following iterative scheme: (1) collect trajectories and measure the dissipated work for each, beginning with a linear initial protocol guess; (2) reconstruct a full free energy landscape from the resulting trajectories using the method developed by Hummer and Szabo (See Equation.~\ref{eq:hummerszabo})~\cite{hummer_free_2001} for adapting the Jarzynski equality to internal molecular coordinates; (3) compute the minimum-dissipation protocol for this landscape using AD, and (4) iterate by repeating steps (1) - (3) with increasingly optimized protocols until the free energy estimate converges (see Figure~\ref{fig:algorithm}). In this paper, we generate trajectories via Brownian dynamics in JAX MD~\cite{schoenholz_jax_2021}, but we emphasize that any experimental or MD source suffices; the only requirement is a recording of the time dependent work (Equation~\ref{eq:work}) used in free energy reconstructions. With our approach, we are able to successfully reconstruct a variety of free energy landscapes, including triple well and asymmetric landscapes, with a marked improvement over the performance of the naive linear protocol. This is relevant for the broad class of biomolecules that feature intermediate, metastable configurations on their energy landscapes~\cite{GuptaWoodsideExperimental2011,bustamante_optical_2021}. Our method is in principle applicable in any regime, from weak to strong protocol strength and slow to fast driving. Additionally, the computational cost of reverse-mode AD scales well with the number of input parameters~\cite{baydin_automatic_2017}; in particular, our method scales linearly in memory with the number of protocol dimensions. Here, we demonstrate the successful use of AD for optimizing 1- and 2-dimensional control protocols across a range of barrier heights. 

The remainder of this paper is organized as follows: in Section~\ref{sec:theory} we review the relevant background: free energy landscape reconstruction following Hummer and Szabo~\cite{hummer_free_2001}, computing optimal protocols for barrier crossing problems, and using automatic differentiation on Brownian dynamics simulations in JAX MD. Section~\ref{sec:algorithm} outlines our iterative method, with simulation details supplied in Appendix~\ref{appendix:sim_details}. Our main results are presented in Section~\ref{sec:results}, and we conclude with a discussion in Section~\ref{sec:discussion}.

\section{Theoretical Background}\label{sec:theory}

\subsection{Free energy landscape reconstructions \label{sec:theory_landscapes}}

%[description of the problem: driven barrier crossing with harmonic trap. Cite experimental examples. Hummer Szabo stuff. Review attempts to maximize accuracy by a. minimizing dissipated work, give definition of work. b. minimizing "Jarzynski error". ]

The biomolecular folding process is generally understood to be a diffusive search over an underlying free energy landscape~\cite{DiffusiveSearch1973}. This folding process is often accelerated with an external harmonic biasing potential, such as in laser optical tweezer (LOT) experiments~\cite{collin_verification_2005,Gebhardt_2010,GuptaWoodsideExperimental2011,engel_reconstructing_2014} in an experimental equivalent of umbrella sampling for rare events~\cite{US1977}. The system is driven over energy barriers, bypassing intrinsically slow-timescale dynamics. In the case of LOT, the ends of the molecule are fixed to polystyrene beads trapped in highly focused laser beams that force a harmonic potential on the end-to-end molecular distance (see Figure~\ref{fig:algorithm}a). As the lasers are moved from separation $z_0$ to separation $z$, the molecule forcibly unfolds. External work $W$ (Equation~\ref{eq:work}) is delivered to the system in the process, which can be used to estimate free energy differences through the Jarzynski equality~\cite{jarzynski_nonequilibrium_1997}:
\begin{equation} 
\label{eq:jarzynski}
F(z) - F(z_0) \equiv \Delta F(z) = -\frac{1}{\beta}\ln{\langle e^{-\beta W} \rangle },
\end{equation}
where $\Delta F(z)$ is the free energy difference between the states where the laser traps are separated by $z$ and $z_0$; $W$ is the external work done by the trap; and $\beta$ is the thermal energy unit, $\frac{1}{k_{B}T}$, with Boltzmann constant $k_{\text{B}}$  and temperature $T$. However, what is often of interest is not only $\Delta F(z)$, but also $\Delta F(q)$, where $q$ is an underlying \emph{molecular coordinate}: the end-to-end extension, for example. Hummer and Szabo adapted the Jarzynski equality for free energies as a function of this \emph{molecular coordinate}, computed using discretely binned nonequilibrium trajectory data~\cite{hummer_free_2001}:
    \begin{equation} \label{eq:hummerszabo}
        \Delta F(\ell) = -\beta^{-1}\log \frac{\sum_{t=1}^{t_f} \frac{\lr{\exp{-\beta W_t}\Theta_{\ell t}}}{\eta_t}}{\sum_{t=1}^{t_f} \frac{\exp -\beta V(\ell, \lambda(t))}{\eta_t}}.
    \end{equation}
Here, $\ell = 1,2,\ldots,K$ denotes a bin along the reaction coordinate. We define $t_f$ to be the final time for the experiment. Here $t=1,2,\ldots,t_f$; we set  $\Theta_{\ell t}$ to be the Heaviside function, non-vanishing only when $x(t)$ is inside bin $\ell$; and $\eta_t = \langle \exp(-\beta W_t)\rangle$ with $W_t$ the external work done up to time $t$ (see Equation~\ref{eq:work}). This equation has been widely used to reconstruct molecular free energy landscapes~\cite{GuptaWoodsideExperimental2011,engel_reconstructing_2014,Ciliberto2017Review}, but suffers from poor convergence, with the number of trajectories $N$ required for accurate free energy estimation proportional to
    \begin{equation} \label{eq:convergence}
    N \sim e^{-\beta \langle W^R_{\text{diss}} \rangle} \text{~\cite{jarzynski_rare_2006,YungerHalpernJarzynski2016}}
    \end{equation}
Here, $\langle W^R_{\text{diss}} \rangle$ is the average nonequilibrium dissipated work  performed in the reverse process $\langle W^R_{\text{diss}} \rangle = \langle W^R \rangle + \Delta F$, computed by moving the trap from $z$ to $z_0$ according to the time-reversed protocol. The dissipated work of this time-reversed trajectory varies monotonically with the forward dissipated work, $\langle W^F_{\text{diss}} \rangle$, which will be the target of this paper for improving convergence of free energy estimators~\cite{jarzynski_rare_2006}.

An alternative optimization objective to $\langle W^R_{\text{diss}} \rangle$ was derived by Gore and Ritort~\cite{gore_bias_2003}: the expected error in Jarzynski-based free energy estimates derived from $N$ trajectories in the near-equilibrium regime (where work distributions can be assumed to be Gaussian):
 \begin{equation} \label{eq:JE}
    \mathrm{error} \sim \frac{1}{N\beta^2} \left\{\langle \exp\left(\beta W^R_{\mathrm{diss}} \right)\rangle - 1 \right\}.
    \end{equation}
Geiger and Dellago~\cite{geiger_optimum_2010} use this objective to find optimal protocols, successfully improving the accuracy of free energy estimation for simulations near-equilibrium; however, this approach struggled with the statistical noise encountered with higher barriers~\cite{geiger_optimum_2010}. Other approaches to improving free energy estimation accuracy have included deriving variants of the Jarzynski equality that display better convergence properties~\cite{LiTu2021LinearNoneq,Whitelam2024PRE}; using bidirectional estimators (the analogue of Bennett's Acceptance Ratio (BAR)~\cite{BAR1976} for nonequilibrium dynamics~\cite{BARVarianceShirts2003})~\cite{ZhongDeWeese2024}; and strategically adding noise to trajectories~\cite{Whitelam2025Noise}. This paper focuses on optimizing the naive unidirectional estimator by optimizing the driving protocol, and leaves a treatment of these alternative estimators with AD for future work.

\subsection{Optimal control in driven barrier crossing}\label{sec:opt_barrier_crossing}

Optimal control protocols were first formulated for stochastic systems such as driven barrier crossing in the context of minimum (work) dissipation protocols~\cite{schmiedl_optimal_2007}. In a pioneering study, Geiger and Dellago~\cite{geiger_optimum_2010} computed optimal one-dimensional (1D) barrier-crossing protocols using gradient descent, with gradients estimated numerically from ensembles of stochastic trajectories. The first general analytical treatment of limited-control (1D) protocols for driving a stochastic system over a bistable landscape was supplied by Sivak and Crooks, who used linear response theory to identify optimal protocols as geodesics in the space of thermodynamic geometry~\cite{sivak_thermodynamic_2016}. Their approach was later validated experimentally, as the near-equilibrium protocols they developed were shown to reduce the external work required to unfold a DNA hairpin using LOT~\cite{tafoya_using_2019}. Blaber and Sivak were the first to treat two-dimensional protocols~\cite{blaber_efficient_2022}, where both the position and the stiffness of a harmonic control potential are varied in time. Using 2D control substantially decreased dissipated work, though their method requires inexact interpolation for intermediate driving speed and strength~\cite{blaber_optimal_2022}. Zhong and DeWeese~\cite{ZhongWorkDissipation2022} later derived optimal 1D barrier-crossing limited control protocols valid arbitrarily far from equilibrium by numerically solving Hamiltonian equations given by optimal control theory.  
Subsequently, Zhong and coworkers presented a method for separately designing forward and reverse driving protocols to minimize the error in bidirectional nonequilibrium free energy estimates, based on numerically minimizing a protocol-dependent importance sampling estimator for work~\cite{ZhongDeWeese2024}. The main bottleneck of these exact approaches lies in poor scaling with dimension: quadratic computational complexity with respect to protocol dimension for the bidirectional protocols%for the escort vector field
~\cite{ZhongDeWeese2024}, and exponential memory scaling with the configuration space dimension for the limited control problems~\cite{ZhongWorkDissipation2022}. One proposed solution is that of Whitelam~\cite{Whitelam_info_erasure_2023}, who solves the related problem of optimal bit erasure (involving driving a stochastic system on a bistable landscape) by training protocols parameterized by a neural network using genetic algorithms. Genetic algorithms have been successfully used to train deep neural networks with millions of parameters~\cite{Galvan2021}, suggesting reasonable scaling with protocol dimension. An alternative approach, which we extend here, is that of Engel \textit{et al.}~\cite{engel_optimal_2023}, who solve for optimal barrier crossing protocols arbitrarily far from equilibrium using AD applied to molecular dynamics (MD) trajectories. The computational cost of AD is linear in the number of protocol parameters, and gradient-based methods are typically more computationally efficient than genetic algorithms, directly using objective gradients to minimize the loss.

We consider a 1D barrier crossing problem as follows: Let $x$ denote a Brownian particle evolving according to a time-dependent energy landscape given by 
\begin{equation}\label{eq:landscape}
    H(x,t) = V_0(x) + V(x,t).
\end{equation}
where $V_0$ is the static, underlying landscape of the form: 
\[V_0 = -\frac 1\beta\mathrm{ln} \ \left[\sum_{i=1}^n \exp\left(-\frac \beta 2  \kappa_i (x - w_i)^2  + \beta E_i\right)\right ],\]
where $n$ denotes the number of barriers. For each well $i=1,2,\ldots,n$, the parameters $\kappa_i$, $w_i$ and $E_i$ denote the curvature, midpoint, and energy, respectively. To steer the particle along this landscape, the Hamiltonian includes a moving harmonic potential of time-varying stiffness $V(x, t) = k_s(t)(x - \xi(t))^2$, which, for example, represents the potential generated by laser beams in LOT experiments. See Appendix~\ref{appendix:sim_details} for exact parameter values used. We calculate the gradients $\nabla_\xi \langle W \rangle$ and $\nabla_{k_s} \langle W \rangle$ to perform a version of stochastic gradient descent via the Adam optimizer~\cite{kingma_adam_2017} in order to find $k_s(t)$ and $\xi(t)$ which minimize $\langle W \rangle$.

%When optimizing protocols with AD, we directly use the gradients of the dissipated work with respect to control parameters to optimize protocols. 
To compute $\nabla_\xi \langle W \rangle$ and $\nabla_{k_s} \langle W \rangle$, we use policy gradients~\cite{engel_optimal_2023,REINFORCE_1992} from the reinforcement learning literature (see also \cite{geiger_optimum_2010}). Equation \ref{eq:reinforce} is known as the REINFORCE gradient~\cite{REINFORCE_1992}.  \begin{align}
\label{eq:reinforce}
\nabla_\theta \mathcal{L}_{\mathrm{DW}}(\theta) &\triangleq \nabla_{\theta} \lr{W(X)} \notag\\
&= \lr{ (\nabla_\theta \ln p(X)) W(X)} + \lr{\nabla_\theta W(X)},
\end{align}
where $X$ denotes a single trajectory in path space, steered for a total switching time $t_f$; $p(X)$ is the probability of realizing a particular trajectory; and work is calculated according to
\begin{equation}
\label{eq:work}
W(X) = \int_0^{t_f} \frac{\partial H\left(X(t), \lambda(t)\right)}{\partial \lambda}\frac {\partial \lambda(t)}{\partial t} dt.
\end{equation}
For the remainder of the paper, we will take dependence on trajectory $X$ to be implicit in any reference to the work W: $W \equiv W(X)$. In general, $W = \Delta F + W_{\mathrm{diss}}$. Since $\Delta F$ is constant with respect to the protocol, minimizing $\langle W \rangle $ is equivalent to minimizing $W_{\mathrm{diss}}$, hence the terminology \emph{minimal-dissipation} protocols. %By keeping track of $p$ for all integration steps of our Brownian trajectories, we can construct 

We use JAX MD to simulate Brownian dynamics~\cite{schoenholz_jax_2021} and AD to collect the gradients $\nabla_\theta W$ and $\nabla_\theta \ln p(X)$. Software which implements AD builds a computational graph and allows for (back)propagation of gradients using the chain rule~\cite{baydin_automatic_2017}. There are two styles of AD, \emph{forward-mode}, which tracks computations and stores the gradient as the computation is being made, and \emph{reverse-mode}, which stores all intermediate gradients, but computes the chain rule from the final computations first. For the purposes of optimal control, the output dimension (e.g. the loss) is typically scalar, and hence by using reverse-mode AD the cost of increasing the dimensionality of control parameters is negligible. In particular, the primary computational limitation is the memory requirement to store the computational graph, which is dominated by the number of simulation timesteps. If the full computational graph is too expensive to store, the memory concerns can be alleviated by checkpointing, at the price of recomputation~\cite{baydin_automatic_2017}.

%For the purposes of free energy landscape reconstruction, a minimal dissipation protocol has no guarantees to minimize the bias of landscapes. Optimizing alternative quantities which contain information of higher order moments of dissipated work can improve free energy estimation \cite{geiger_optimum_2010}. Optimization of such higher-moment objectives were attempted, but the optimization landscape proved too noisy to achieve converged results (See Appendix~\ref{appendix:other_loss}).

%In addition, impressive iterative protocol optimization procedures have been developed by minimizing an importance-sampled objective~\cite{ZhongDeWeese2024}. Inspired by such techniques, we propose a complementary approach in which JAX-MD simulations can be used alongside experimental techniques via an iterative procedure to construct entire free energy profiles. These computational gains can be used in collaboration with experiment to resolve noisy landscapes, possibly far out of equilibrium. For example, many DNA hairpins have been shown to have several intermediate states, which often cause high errors when computing free energy profiles~\cite{GuptaWoodsideExperimental2011}.

\section{Iterative Algorithm}\label{sec:algorithm}
Our proposed iterative algorithm begins with a linear initial protocol for $k_s^0(t)$ and $\xi^0(t)$ and iteratively updates the protocol based on experimental or simulated information from the underlying system. The process is as follows:
\begin{enumerate}
    \item Perform nonequilibrium driven barrier crossing simulations or LOT experiments using stiffness and trap position protocols $k_s^i(t)$ and $\xi^i(t)$, collecting trajectories and measuring the dissipated work for each (without the need to calculate gradients);
    \item Reconstruct a candidate full free energy landscape from the resulting trajectories using the method developed by Hummer and Szabo~\cite{hummer_free_2001} for adapting the Jarzynski equality to internal molecular coordinates (Equation \ref{eq:hummerszabo}); 
    \item Compute the minimum-dissipation protocols for this landscape using AD on 1D Brownian dynamics simulations in JAX MD as described in section \ref{sec:opt_barrier_crossing}, which become $k_s^{i+1}(t)$ and $\xi^{i+1}(t)$;
    \item Iterate by repeating steps (1) - (3) until convergence.
\end{enumerate}
Our algorithm is summarized in Algorithm \ref{alg:iterative-algo} and depicted in Figure~\ref{fig:algorithm}. This iterative method allows for the optimization of protocols in a tractable configuration space, even if sampling the ambient experimental space is expensive (e.g., high-dimensional molecular dynamics or real LOT experiments). In particular, the low dimensionality and simple dynamics of the Brownian particle allow for rapid iteration. 

\begin{algorithm}
\DontPrintSemicolon
\caption{Iterative procedure for reconstructing unknown free energy landscapes using (Eq. \ref{eq:hummerszabo}) and AD-optimized protocols.}
\label{alg:iterative-algo}
\KwIn{$\mathrm{oracle}(\lambda): \R \to \R$, $\lambda_{\mathrm{init}} \in \R^{t_f}$}
\Parameter{$\mathrm{numBatches}$, $\mathrm{numEpochs}$}
\KwOut{Optimized Free energy landscape}
        $\lambda \gets \lambda_{\mathrm{init}}$
        // Run reconstruction algorithm for $\mathrm{numEpochs}$ or until convergence
        
        \For{$i\gets 0$ \KwTo $\mathrm{numEpochs}$ }{ 
    	\For{$n\gets 0$ \KwTo N}{
                $\mathrm{trajectories}[n], \mathrm{works}[n]  \gets \mathrm{oracle}_{\lambda}$
                }

            $\mathrm{newLandscape} \gets \texttt{HummerSzabo}(\mathrm{trajectories}, \mathrm{works}, \lambda)$

            $\lambda \gets \texttt{Optimize}(\mathrm{newLandscape}, \lambda)$
        
        }
	\Return $\mathrm{newLandscape}$
\end{algorithm}

In order to quantify the performance of a protocol in reconstructing a free energy landscape, we define the \emph{landscape bias} as the maximum difference between the estimated and true landscape free energy. The maximum is taken over the relevant regions of the reaction coordinate (between two endpoint wells), and after aligning the first free energy minima. If $\Delta\hat{\mathbf{F}} = \{\Delta F(\ell)\}_{\ell = 1}^K$ is our reconstructed, estimated landscape using $K$ bins in the reaction coordinate, the \emph{landscape bias} is 
\begin{equation}
\label{eq:bias}
\bias = \max_{\ell \in \{1,2,\ldots, K\}} \| \Delta F(\ell) - \Delta \hat F(\ell)\|.
\end{equation}

\begin{figure}
    \centering
    \vspace{1em}
    \subfloat[Left panel, 1\,\si{\milli\second} simulations. Right panel, 100\,\si{\micro\second} simulations.]{\includegraphics[width=0.8\linewidth]{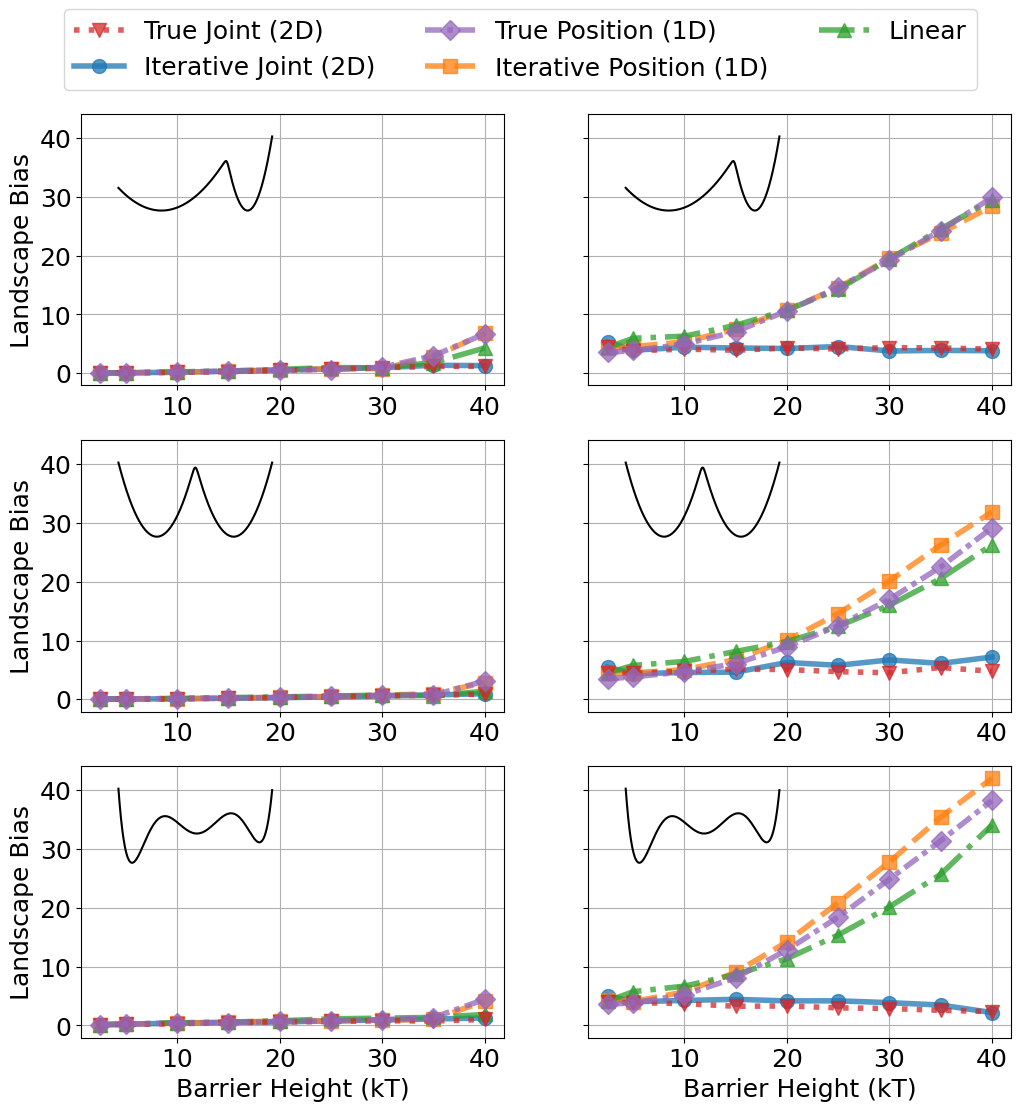}}\\
    
    \subfloat[1\,\si{\milli\second} simulations, up to 100$\kT$ barrier, extending bottom-left plot of (a).]{
    \includegraphics[width=0.6\linewidth]{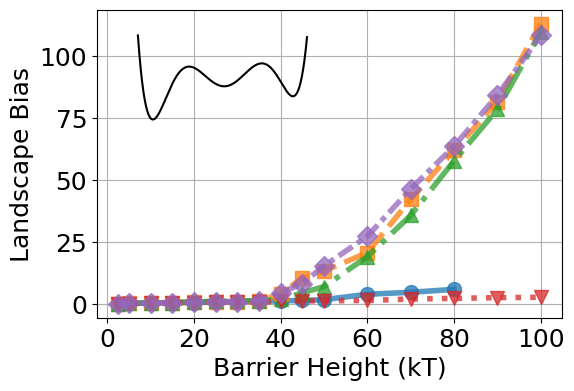}}
    \caption{Performance of the iterative landscape reconstruction scheme as quantified by the landscape bias, Equation \ref{eq:bias}, for asymmetric, bistable symmetric, and triple well landscapes; various barrier heights; and for slower (left panel, (a)) and faster (right panel, (a)) driving. The results of using the iterative procedure for 1D (orange) and 2D (blue) control are compared to using the \textit{a priori} knowledge of the landscape directly for 1D (purple) and 2D (red) control. Also shown is the landscape bias resulting from using a naive linear protocol (green). Landscapes were generated with 1000 trajectories. For 1D control, all simulations used a default $k_s = 0.4$\,\si{\pico\newton/\nano\meter} throughout the entire simulation.}
    \label{fig:bias-barrier}
\end{figure}

\begin{figure}
    \centering
    \includegraphics[width=0.8\linewidth]{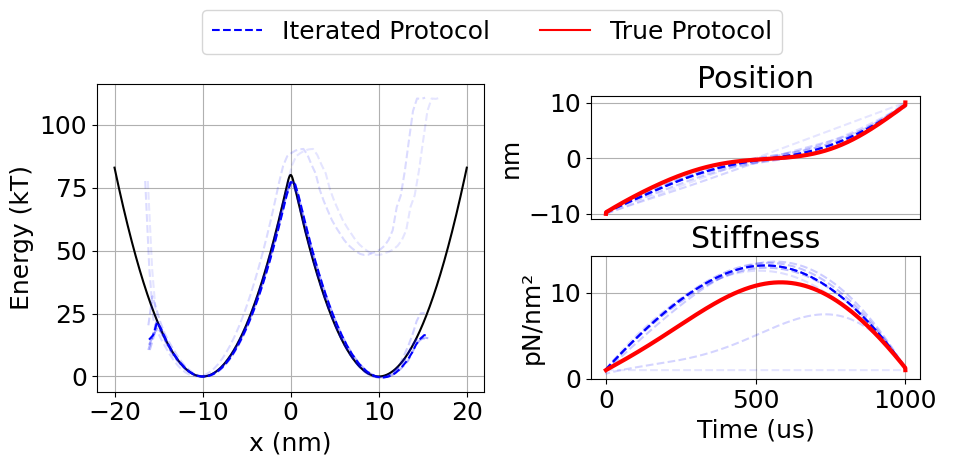}
    \includegraphics[width=0.8\linewidth]{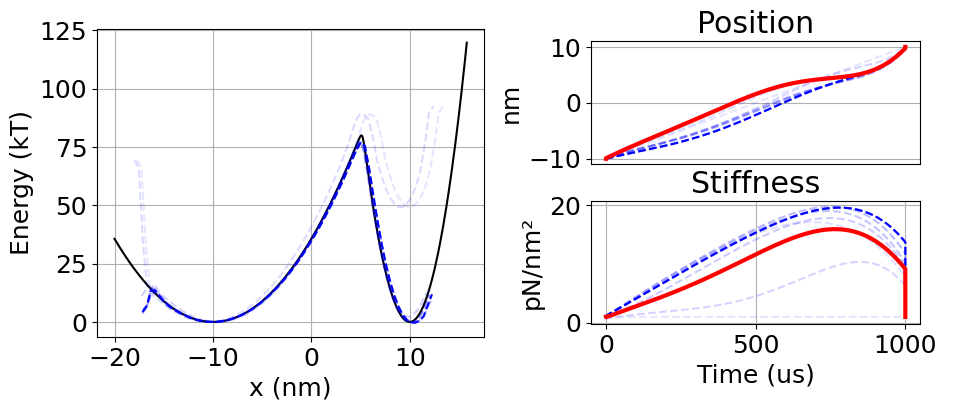}
    \includegraphics[width=0.8\linewidth]{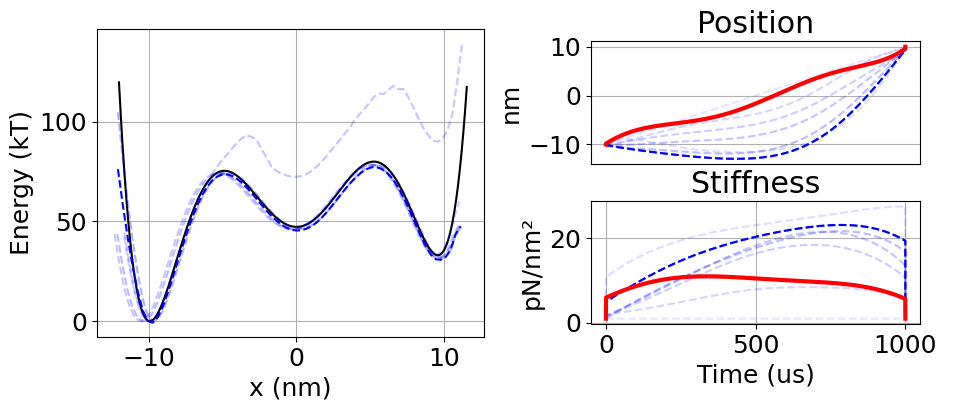}
    \caption{Iterative scheme of 2D protocol landscape reconstructions and optimized protocols for 1\,\si{\milli\second} simulations with 80$\kT$ barrier height. Each landscape had 10 total iterations, with the darkest line representing the protocol for the final iteration. 2D protocols optimized with the iterated scheme recovers near-perfect landscape reconstruction performance of 2D protocols optimized with gradients of the true underlying landscape, despite differing in form from the protocols that used knowledge of the `true' landscape (shown in red), indicating degeneracy in the loss landscape.}
    \label{fig:it-opt-protocols}
\end{figure}

\begin{figure}
    \centering
    \includegraphics[width=1\linewidth]{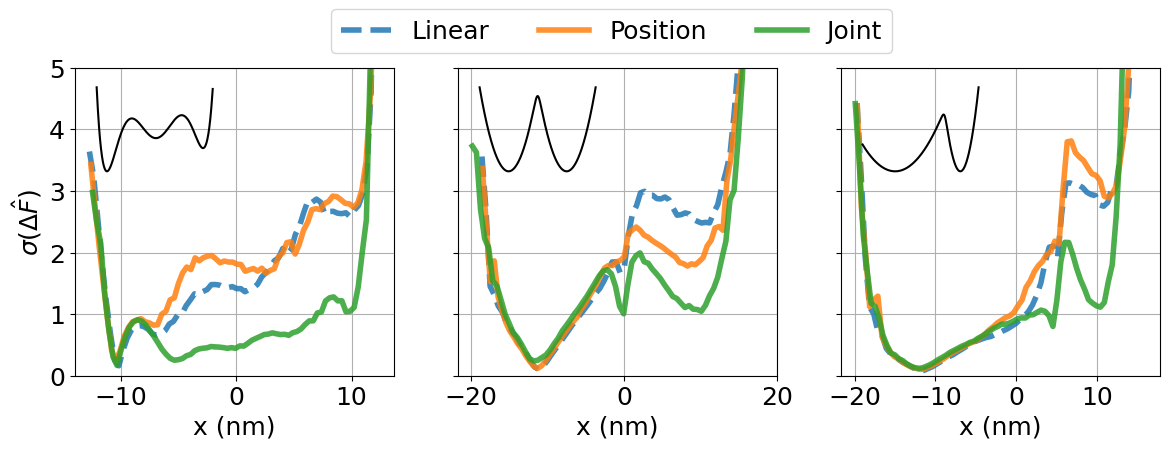}

        \caption{Standard deviation of free energy landscape reconstructions. Individual landscapes are constructed with $300\si{\micro\second}$ simulations and an $25\kT$ barrier height, with 1000 trajectories per landscape. For every protocol type we generated 100 independent landscapes; $\sigma\left(\Delta\hat{F}\right)$ denotes the empirical standard deviation at each position bin, computed across those landscapes. }
    \label{fig:std-reduction}
\end{figure}

\begin{figure}[hb]
    \centering
    \includegraphics[width=\linewidth]{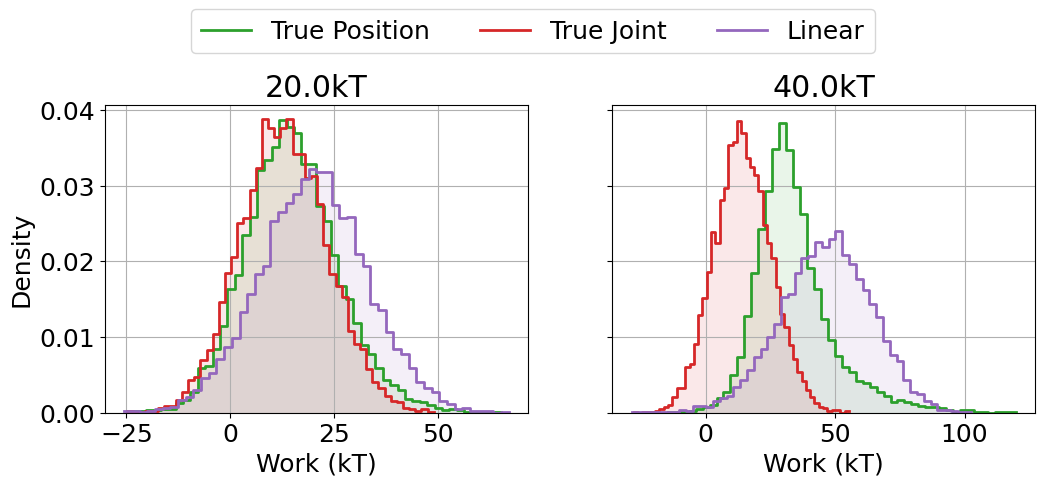}
    \caption{Histograms of work distributions for 300\,\si{\micro\second} simulations at near-equilibrium 20$\kT$ (left) and far-from-equilibrium 40$\kT$ (right) barrier heights. Far-from equilibrium, AD optimized 1D protocols (True Position, green) increase skewness, despite decreasing average work. The increased positive skewness decreases landscape reconstruction accuracy, explaining the poor performance of 1D protocols (True Position) in Fig.~\ref{fig:bias-barrier}.}
    \label{fig:work-distributions}
\end{figure}

\section{Results}\label{sec:results}

%%%% 

%[We show in Appendix~\ref{appendix:reproduce_protocols} that AD can additionally reproduce 2D linear response derived by Blaber \emph{et al.}~\cite{blaber_efficient_2022}, and the non-monotonic shape of linear biasing protocols discovered by Zhong \emph{et al.}~\cite{ZhongDeWeese2024, ZhongWorkDissipation2022}. protocol degeneracy ]

We first validate our AD-based approach to barrier-crossing optimal control by using it to successfully recover the non-monotonic optimal protocol seen by Zhong and DeWeese~\cite{ZhongWorkDissipation2022} for a linear biasing potential (See Appendix~\ref{appendix:reproduce_protocols}).
%Performance of AD optimized protocols for both 1D and 2D control is presented in Fig.~\ref{fig:bias-barrier}.

Next, we demonstrate the ability of 2D AD optimized protocols to successfully reconstruct unknown landscapes by applying our iterative algorithm (Algorithm~\ref{alg:iterative-algo}, Fig.~\ref{fig:algorithm}) to three different landscape types: bistable symmetric, triple well (representing, for example, a biomolecule with an intermediate on its folding pathway), and bistable asymmetric. All simulations have position protocol endpoints specified at $x = -10~\si{\nano\meter}$ and $x=10~\si{\nano\meter}$, which correspond to the first and final well respectively. Barrier heights, measured in units of $\si{\kT}$, are measured from the bottom of the initial well to the highest barrier between the start and end points. Higher barriers,  sharper potential curvatures, potentials with relevant intermediate states, and entropic traps all increase the natural relaxation timescale of the system, requiring a stronger nonequilibrium drive to cross these barriers at a fixed timescale. We denote AD protocols optimized using JAX MD simulations of a particle with the \emph{true} underlying landscape (i.e. using \textit{a priori} knowledge of the landscape) as either \emph{True Joint} (2D) or \emph{True Position} (1D) protocol, and we emphasize these protocols are what ideally the iterative scheme converges to, as both methods use AD with the same loss to find optimal protocols. For all 1D control experiments, we use relatively weak driving at $k_s = 0.4$\si{\pico\newton/\nano\meter}, representative of LOT experiments. Complete simulation and landscape details are provided in Appendix~\ref{appendix:sim_details}.

In Figure~\ref{fig:bias-barrier}, we simultaneously present the impressive ability of 2D protocols (denoted \emph{joint} protocols in the figures) to reconstruct all three types of landscapes far from equilibrium, as well the success of the iterative scheme in recovering the same results with minimal error. In the near-equilibrium regime, we are able to recover underlying landscapes nearly perfectly,  even when using 1D protocols, which is expected from literature~\cite{schmiedl_optimal_2007,geiger_optimum_2010,engel_optimal_2023}. For slow driving (left panel in Figure~\ref{fig:bias-barrier}(a) and (b)), the iterative scheme recovers the landscape with low bias for both 1D and 2D protocols up to $\sim40 \kT$ barrier heights, beyond which 2D protocols are needed to accurately reconstruct the landscape. The iterative method performs as well as the ``true'' optimized protocols save for barriers $\geq 90 \kT$ on the triple well landscape, where the particle did not successfully cross the barriers. For fast driving, the iterative method performs as well as direct protocol optimization on the true underlying landscape, both of which are able to recover landscapes out to a substantially far-from-equilibrium regime. Such performance of 2D control protocols is consistent with the findings of Blaber and Sivak~\cite{blaber_efficient_2022}.  Samples of the recovered landscapes, along with the converged protocols, are shown in Figure ~\ref{fig:it-opt-protocols}. We find that even very far from equilibrium -- at $100\kT$ barrier heights with $t_f = 1$\si{\milli\second} -- convergence occurred within 1-4 iterations of Algorithm~\ref{alg:iterative-algo}, with the exception of poor convergence in triple well experiments which began around barrier heights of $90\kT$ due to lack of full barrier crossings.% Note the degeneracy in solution space: the protocols from the iterative scheme often differ from those that result from a straightforward protocol optimization on the ``true'', underlying landscape, but perform similarly well in reconstructing the landscape. Such a phenomenon was studied by Gingrich \emph{et al.} in~\cite{GingrichDegeneracy2016}. This underscores the non-convexity of the loss landscape for barrier crossing problems and highlights the fact that while our method does not produce guaranteed \textit{global} optima, it can recover multiple local optima with similar performance. 

%Empirically, we find for high barriers, the iterative scheme seems to converge to a protocol which does not prioritize barrier crossings. This seems to successfully results in low dissipation protocols, but is problematic for free energy estimation and reconstruction (See Appendix~\ref{appendix:protocol-kl}). 

We note that the method continues to perform well remarkably far from equilibrium: for landscape bistable barriers of 100~\si{\kT}, we are still able to recover the landscape with 90\% accuracy (i.e. the bias is $\leq 10\%$ for a 100~\si{\kT} barrier) for simulations lasting $t_f$ = 1\si{\milli\second}; see Figure~\ref{fig:it-opt-protocols}. For faster driving, with simulation lengths $t_f$ = 300\si{\micro\second} and $t_f$ = 100\si{\micro\second}, we can achieve 10-20\% error for 40~\si{\kT} barriers, with our method still outperforming the naive protocol (linear trap motion and constant stiffness) by roughly an order of magnitude.

Not only can the iterative scheme alongside 2D control recover the true landscape on average; it substantially reduces the variance in landscape reconstructions. As shown in Figure~\ref{fig:std-reduction}, even with 1000 trajectories--a number already quite large for experimental feasibility--the standard deviation of landscape reconstructions using 1D control remains prohibitively high far from equilibrium, reaching up to 15\% of the barrier height during relevant portions of the landscape. In contrast, 2D control reduces standard deviation by around 3-fold at the end points.

In all cases, employing an extra protocol dimension--trap stiffness in addition to trap position--improves free energy landscape reconstructions, as expected~\cite{blaber_efficient_2022}. The relative performance of 2D compared to 1D steering seems to be primarily due to the ability to stiffen the trap substantially near barrier crossing--where the optimal stiffness is on the order of $\sim20$\si{\pico\newton/\nano\meter} for 100$\kT$ barriers (se Fig.~\ref{fig:it-opt-protocols}). Empirically, we find this stiff trap will keep work distributions close to the equilibrated Boltzmann distributions, despite the nonequilibrium driving; see Appendix~\ref{appendix:protocol-kl} for further details. 

When the system is far from equilibrium under relatively weak driving, one-dimensional optimal control can underperform a simple linear protocol. Cases of minimal-dissipation protocols yielding degraded free-energy estimates compared to naive steering have been reported previously~\cite{GingrichDegeneracy2016,DavieCompression2014,blaber_skewed_2020}. We explain the discrepancy in our example of barrier crossing by comparing the work probability distributions (Fig.~\ref{fig:work-distributions}) of 1D optimal protocols. Although we see reduction of average work upon optimization, we find that as we get farther from the near-equilibrium regime, work distributions become increasingly non-Gaussian. In particular, 1D minimal-dissipation protocols reduce dissipation at the expense of a higher positive \emph{skewness} (computed as $\left\langle(W-\langle W \rangle)^3\right\rangle/\sigma_W^3$) of the distribution compared to a naive linear protocol. Since Jarzynski's equality depends on maximizing the number of trials observed in the low-dissipation tail of the work distribution, this causes the landscape reconstructions to degrade in quality relative to the linear protocol ones. 
An alternative objective function is the expression for Jarzynski error given by Equation~\ref{eq:JE}. We found that in the far-from-equilibrium regime, while these optimizations indeed increased the negative skewness of the work distributions, the reconstructions were inferior to those that used average dissipated work as an objective function; see Appendix~\ref{appendix:other_loss} for details. This is presumably because the assumption of Gaussian work distributions that underlies the derivation of Equation~\ref{eq:JE} breaks down in the far-from-equilibrium regime, and higher-order cumulants become important in the calculation of the variance of the Jarzynski estimator~\cite{gore_bias_2003,bias-variance}.

\section{Discussion}\label{sec:discussion}

%%% From here it is drafts
%%% Need to talk about optimizing loss functions (might be better in discussion?

%%% Need to explain why one control param not enough (work distributions)
Our work introduces an iterative method to probe energy landscapes and perform automatic differentiation (AD) enabled optimization when the full energy function is unknown (e.g., in experiments). We successfully reproduce and extend Blaber and Sivak’s efficient two-dimensional control for barrier crossing using AD, and find that our method remains effective even when the initial steering is far from equilibrium. Importantly, the methods we introduce use AD to optimize protocols, which scales well in protocol dimension, enabling the use and optimization of multi-dimensional protocols and multiple steering instruments.

In general, there is no guarantee that the protocol AD converges on is \textit{globally} optimal (though we can expect it to be locally optimal). However, the AD protocols perform well for the low-dimensional examples considered here. Of note is the fact that almost perfect reconstructions are achieved by the iteratively-derived protocols in Figure~\ref{fig:it-opt-protocols} despite the iterative method converging on \emph{different} protocols to those derived using the ``true'' underlying landscape. This suggests degeneracy in the loss landscape, which has been noted elsewhere in the context of optimal control~\cite{GingrichDegeneracy2016,engel_optimal_2023}. To verify this was not an issue of convergence for the iterative scheme, we ran further gradient descent with the true landscape as the underlying $V_0$ and starting from the iterative method's result as an initial guess and found the protocol did not change substantially.

A natural next step is to test our iterative scheme in higher dimensions. Although two-dimensional control performs well, a single harmonic trap is unlikely to keep correlated, high-dimensional systems (e.g., protein–ligand) near equilibrium. A practical route forward is to reduce dimensionality to key reaction coordinates and apply a control schedule on each coordinate. For example, each coordinate could be controlled by a separate trap, whereby the schedule for each trap will be optimized in a scalable fashion with AD. We expect the primary bottleneck to be storing the computational graph for such longer timescale simulations.

While we have shown how effective 2D control can be with minimum dissipation optimizations, 1D protocols optimized for minimum dissipation can struggle to reconstruct landscapes accurately. In addition, even for 2D protocols, the required trap stiffnesses ($\sim$20\si{\pico\newton\per\nano\meter}) may be experimentally unfeasible to implement. Future work can exploit one of the major benefits of AD: the ease with which one can define and experiment with objectives (loss functions) to minimize. For instance, one can calibrate the loss to the particular experimental set-up, adding penalties depending on the constraints of the experiment (e.g. a regularizer to penalize high coefficients on stiffness protocols).

In Appendix~\ref{appendix:other_loss}, we show some results of using Jarzynski Error (Equation~\ref{eq:JE}) as an objective function. We find that the assumption of Gaussianity of work distributions underlying Equation~\ref{eq:JE} is problematic for far-from-equilibrium optimizations (see the green, ``True Position'' distribution in Fig~\ref{fig:work-distributions}(b), for example), where higher-order cumulants become important in the calculation of variance of the Jarzynski estimator~\cite{gore_bias_2003,bias-variance}. One possible way forward is to use the variance or negative skewness of work distributions as loss functions, and this is a direction to be explored in future work.

In this vein, a promising direction would be to minimize the variance of bidirectional estimators, derived for non-equilibrium dynamics by Shirts \emph{et al.} in~\cite{BARVarianceShirts2003}. Unfortunately, direct minimization using AD will require tackling the difficult problem of performing AD through the iterative optimization required to compute the free energy necessary in the BAR equations, and hence alternative methods must be explored. Another exciting direction for future work is in the optimization of ``escort'' vector fields, or counteradiabatic driving forces with AD (as opposed to limited control)~\cite{SuriJarzynskiEscort2008}, as was recently made tractable far from equilibrium by Zhong \textit{et al.}~\cite{ZhongDeWeese2024}.  As noted in \cite{ZhongDeWeese2024}, full-control optimal escort vector fields act as perfect stochastic normalizing flows~\cite{snfNoe2020} in the machine learning literature, and it is of great interest to further integrate these fields in order to explore both computationally efficient and exact optimization algorithms.

\section*{Acknowledgments}

This research was supported by the Office of Naval Research through ONR N00014-17-1-3029 and N00014-17-1-3029 and the NSF AI Institute of Dynamic Systems (\#2112085).
M.C.E. thanks Schmidt Futures in partnership with The Rhodes Trust for partially funding this work and acknowledges the support of the Natural Sciences and Engineering Research Council of Canada (NSERC), funding reference number RGPIN-2024-06144.
Z.A. acknowledges support from the Caltech Ph 11 Research Fellowship (2022).
\bibliographystyle{apsrev4-2}
\bibliography{refs.bib} 
%\end{document}

\clearpage
\appendix

\section{Simulation details}\label{appendix:sim_details}
We consider a system governed by overdamped Langevin Dynamics (Brownian motion) with the following limited control, time-dependent potential.
\begin{equation}
V(x,t) = V_0(x) + \frac 12 k_s(t)(x - \xi(t))^2,
\end{equation}
Trap stiffness $k_s$ and trap position $\xi$ denote our control parameters, and the underlying potential $V_0$ is unknown \textit{a priori}. The equation of motion is
\begin{equation}\label{eq:langevin}
    m\gamma x'(t) = -\nabla_x V_t - B_t \sqrt{2\kT},
\end{equation}
where $B_t$ is a Wiener process and represents random interactions with a thermal bath: $\langle B_t B_{t'}\rangle = \delta(t-t') $ and $\langle B_t\rangle = 0$. 
For modeling biomolecules in particular, $B_t$ models random fluctuations that occur from interactions with solvent. For the main text, we used parameter values as follows: temperature $T= \beta^{-1} = 4.183 \kT$ at $303K$, mass $m = 10^{-17}$~\si{\gram}, and friction coefficient $\gamma = \left(\beta \cdot (0.44 \cdot 10^6) \cdot m\right)^{-1} = 9.5069\cdot 10^{11}$ \si{s^{-1}}. For constant stiffness protocols, we defaulted to $0.4$ \si{pN}/\si{nm}, in the typical range for LOT experiments. The naive, linear protocols used in the text correspond to a constant trap speed and a constant stiffness of $0.4$ \si{pN}/\si{nm}. All parameters are also available on our \href{https://github.com/RubberNoodles/barrier_crossing}{GitHub}. We use JAX MD to simulate and collect gradients via automatic differentiation.

To estimate the free energy, we require measurements of the thermodynamic work exerted by the control system, defined in Equation~\ref{eq:work}. Numerically, we alternate between performing control parameter updates (moves in $\lambda$) and molecular coordinate updates (moves in $x$). Following Crooks~\cite{Crooks1998}, we consider the work in each step to be accrued upon changing $\lambda$ and the heat to be accrued upon changing $x$. Thus, we sum over changes in the Hamiltonian for each $\lambda$ parameter update, holding $x = X(t)$ constant:
\begin{equation*}
    W(X) = \sum_{t=0}^{t_f-1} H(X(t), \lambda_{t+1}) - H(X(t), \lambda_t).
\end{equation*}

\subsection{Landscapes}

\begin{figure}
    \centering
    \includegraphics[width=0.5\textwidth]{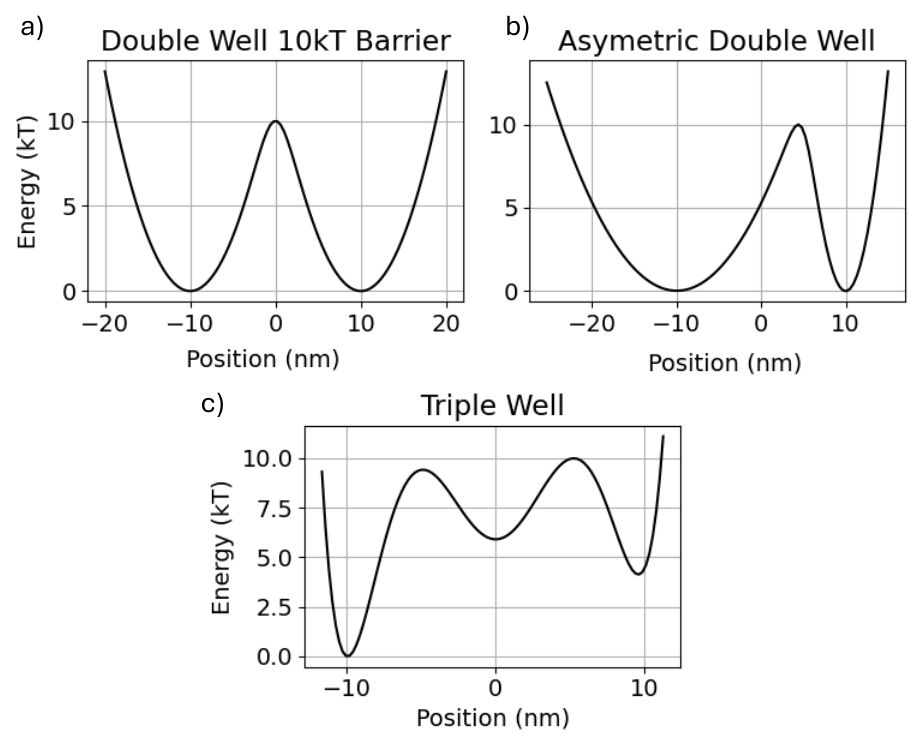}
    \caption{Potential Energy Landscapes. (a) Standard bistable potential. (b) Asymmetric bistable potential. (c) Triple well potential, modeling intermediate states.}
     \label{fig:landscapes}
\end{figure}

We used a general family of functional forms to model a variety of energy landscapes (see Figure~\ref{fig:landscapes}), generalizing previous work on bistable potentials~\cite{geiger_optimum_2010, sivak_thermodynamic_2016}. We consider a potential energy surface with $n$ wells along a reaction coordinate $x$. Label these wells $w_1,w_2,\ldots,w_n$ with energies $E_1,E_2,\ldots,E_n$ and curvature $\kappa_1, \kappa_2,\ldots,\kappa_n$. This leads to the family of potentials we use for the simulations in the main text.
\begin{equation} \label{eq:sc landscapes}
    V_0(x) = -\frac 1\beta\mathrm{ln} \ \left[\sum_{i=1}^n \exp\left(-\frac \beta 2  \kappa_i (x - w_i)^2  + \beta E_i\right)\right ].
\end{equation}
In order to reproduce results from the literature in Appendix~\ref{appendix:reproduce_protocols}, we use the double well potential $V_0 = (x^2-1)^2$.
\begin{comment}as well as the potential used by Geiger and Dellago in~\cite{geiger_optimum_2010}:
\begin{align}
    V_{\varepsilon,\sigma}(x,t) &=\frac k2 (x - \lambda_\theta(t))^2 +  \varepsilon(1-(x/\sigma - 1)^2)^2 \notag\\
    &\qquad\qquad\qquad+\frac{\varepsilon}{2}(x/\sigma - 1)
    \label{eq:gd_functional}
\end{align}
\end{comment}
\label{appendix:landscapes}

\section{Optimal protocols and distance from equilibrium}
\label{appendix:protocol-kl}
There has been substantial work done to understand the connections between optimal control, thermodynamic geometry, and optimal transport; see for example~\cite{blaber_efficient_2022,OTProesmans2020,OTAurell2011,Zhong2024ThermoGeomOC}. As mentioned, we find that 2D control keeps the ensemble distribution of the brownian particle close to that of the equilibrium Boltzmann distribution. To quantify this, we calculate the Kullback-Leibler (KL) divergence between the empirical distribution of positions and the Boltzmann (equilibrium) distribution at the given timestep: $p(x) = e^{-\beta H_\lambda(x)}/Z_\lambda$, with $p(x)$ the probability of observing a given position coordinate, $H_\lambda(x)$ the Hamiltonian under protocol $\lambda$, and $Z_\lambda$ the partition function. The KL divergence between two probability distributions $p,q$ is defined as
$$\mathrm{KL}(p \| q) = \left\langle \ln\frac pq \right\rangle_p.$$
Crucially, this KL divergence vanishes when $p=q$, which will be the case when the distribution is exactly in equilibrium. The KL divergence at the beginning of our simulations will be 0 by construction, up to numerical error, then become nonzero as the non-equilibrium protocol perturbs the system.
\begin{figure*}
    \centering
    \includegraphics[width=0.45\textwidth]
    {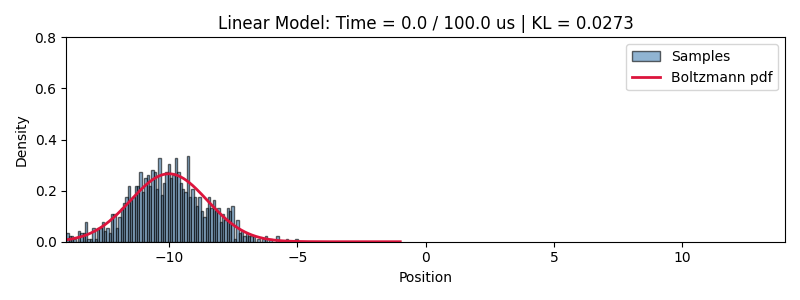}
    \includegraphics[width=0.45\textwidth]
    {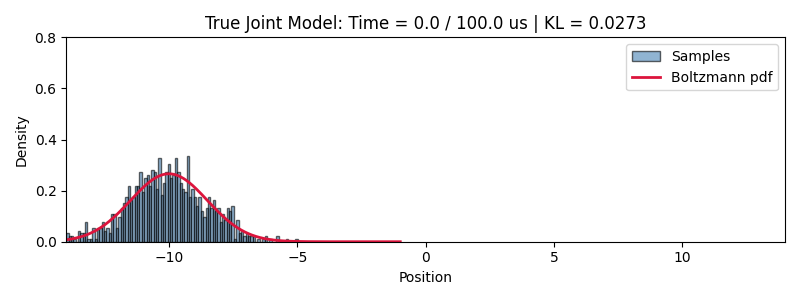}
    \includegraphics[width=0.45\textwidth]
    {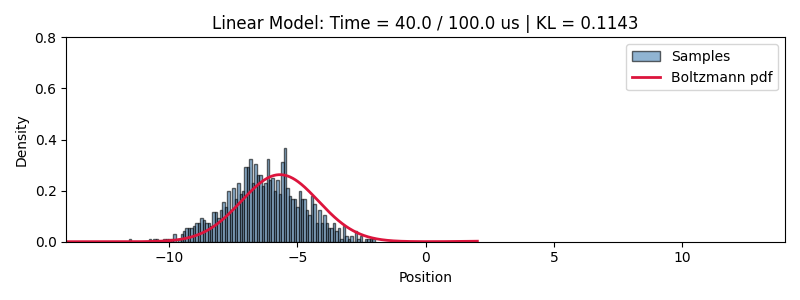}
    \includegraphics[width=0.45\textwidth]
    {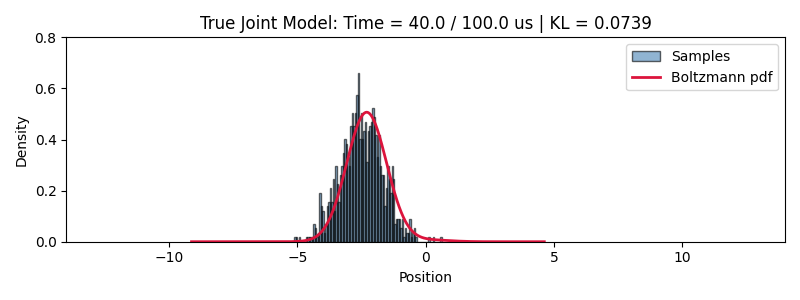}
    \includegraphics[width=0.45\textwidth]
    {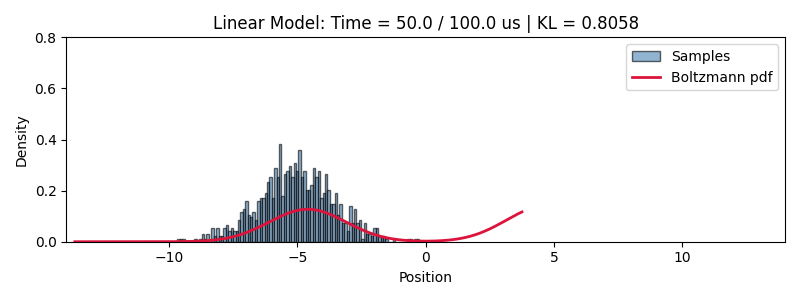}
    \includegraphics[width=0.45\textwidth]
    {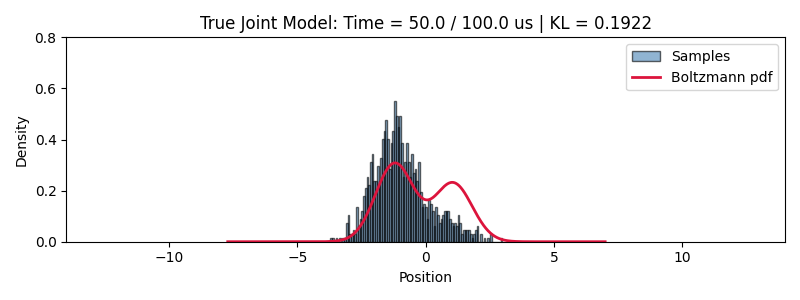}
    \includegraphics[width=0.45\textwidth]
    {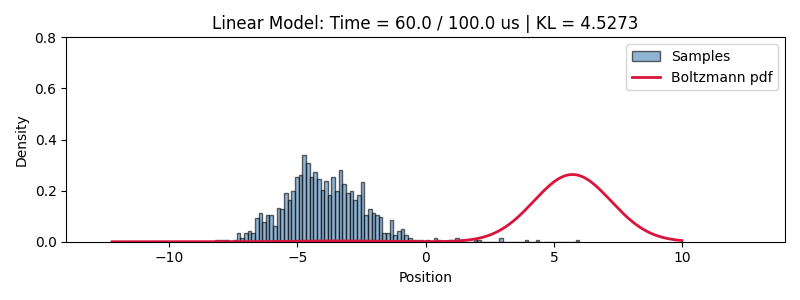}
    \includegraphics[width=0.45\textwidth]
    {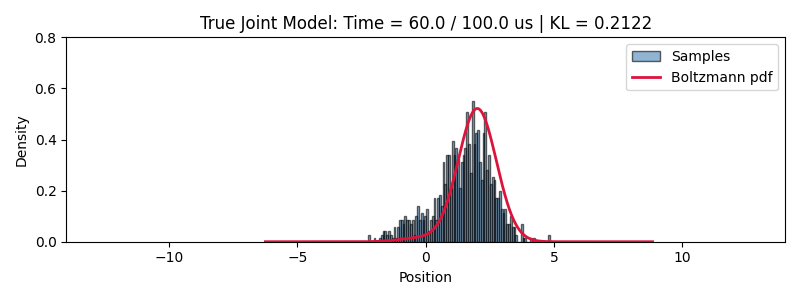}
    \includegraphics[width=0.45\textwidth]
    {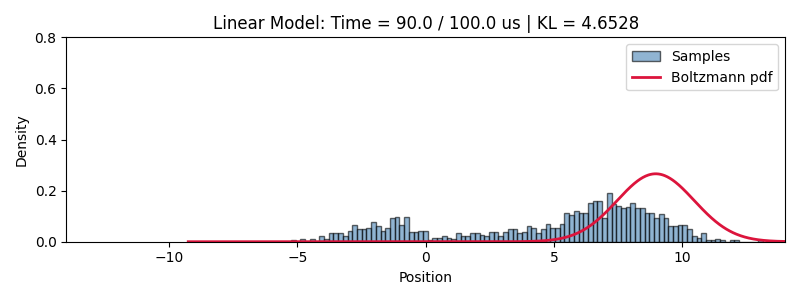}
    \includegraphics[width=0.45\textwidth]
    {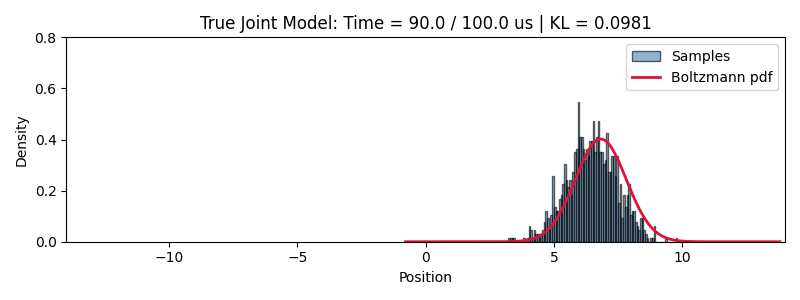}
    \includegraphics[width=0.45\textwidth]
    {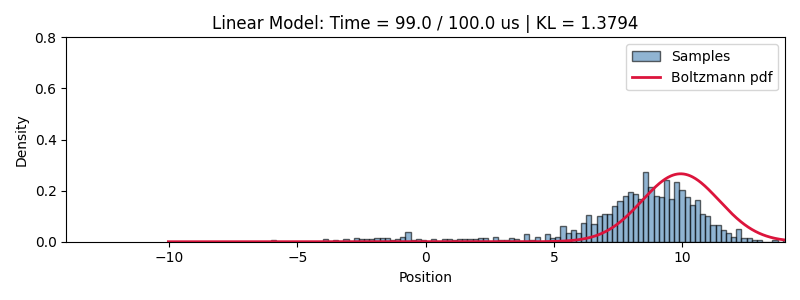}
    \includegraphics[width=0.45\textwidth]
    {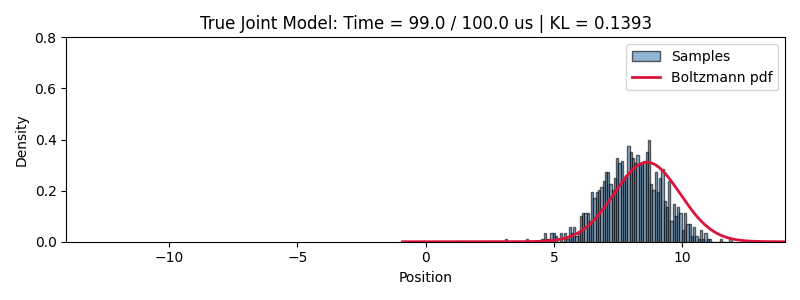}
    \caption{Empirical divergence from equilibrium for linear (left panel) and optimal (right panel) steering for 2D control on an underlying bistable symmetric potential. KL value represents the Kullback-Leibler divergence between an empirical histogram from 1000 trajectories and the true Boltzmann distribution calculated numerically from the energy function. Our results corroborate those of reference \cite{blaber_efficient_2022}.}
    \label{fig:kl-evolution}
\end{figure*}
Reference \cite{blaber_efficient_2022} found that the optimal 2D minimal-dissipation protocol with a harmonic spring is able to keep the intermediate probability distributions close to equilibrium, and we here corroborate their results using the optimal protocols derived from our iterative method. The potential we use for the dynamics is the double-well shown in Figure~\ref{fig:landscapes}, top left and defined by Equation~\ref{eq:sc landscapes}.

\begin{comment}
This toy example is particularly helpful, as we can use optimal control theory to see how limited control "falls short" of full control in our different regimes. We stress that full control is not physically realizable and thus is only used here as an inspiration to improve protocols. We expect this to be particularly important in regimes with many degrees of freedom. 

For driving according to a time-dependent potential $V(x,t)$, we define the probability distribution/phase-space density corresponding to the ensemble at time $t$ as $p_t(x)$. Given we are running with overdamped Langevin dynamics (see Equation~\ref{eq:langevin}), the density evolves according to the Fokker-Planck Equation.
$$\partial_t p = \mathcal{L}_\lambda p$$ 
Where $\lambda$ is the time varying control parameter and $\mathcal{L}$ is the corresponding infinitesimal generator defined as follows.
$$L_\lambda p = \Delta p + \nabla \cdot (p\nabla V_\lambda).$$
\end{comment}

The quality of non-equilibrium free energy estimates is closely tied to the extent to which the evolving non-equilibrium distributions ``lag'' behind the equilibrium distributions~\cite{SuriJarzynskiEscort2008}. Figure~\ref{fig:kl-evolution} shows how our iteratively derived 2D optimal control protocol ensures that there is very little lag in the distribution. Empirically we see that under the linear control protocol, the distributions can't ``mix'' between the wells rapidly enough relative to the protocol speed, while the optimal 2D protocol features a slower protocol speed relative to mixing time, allowing the distribution to remain nearer to equilibrium. As a result, 2D control is able to perform substantially better than linear control, as expected.

% WIP from here. Go through optimal control theoryLet $\mathcal{V}$ consist of the collection of velocity fields $v(x,t)$. 

% Then do some numerical simulations on the equaiton derived in sivak blaber

% It would be best to put these as some kind of line in the appendix fig 7.

\section{Benchmarking AD for linear driving}
 Engel \textit{et al.} previously showed AD can reproduce optimal protocols from linear response theory in near-equilibrium regimes~\cite{engel_optimal_2023}. Here, we validate our method against the work of Zhong \textit{et al.}~\cite{ZhongWorkDissipation2022}, who find a surprising non-monotonicity in minimal-dissipation protocols when the driving potential is a linear function of position~\cite{Zhong2024ThermoGeomOC,ZhongWorkDissipation2022}. To ensure that our algorithm captures this behaviour, we use AD to derive an optimized protocol for time-varying linear biasing. Note that ``linear'' in this case refers to the degree of the contribution $\lambda_\theta(t)$ to the overall potential and not to the shape of $\lambda_\theta(t)$:
\begin{equation}
\label{eq:linear_biasing}
V(x,t) = \frac 14 (x^2 - 1)^2  - x\lambda_\theta(t)
\end{equation}
Here, $\lambda_\theta(t)$ is a time-varying protocol, and we minimize work as our objective (see Equation~\ref{eq:work}).

\begin{figure}
    \centering
    \includegraphics[width=0.5\textwidth]{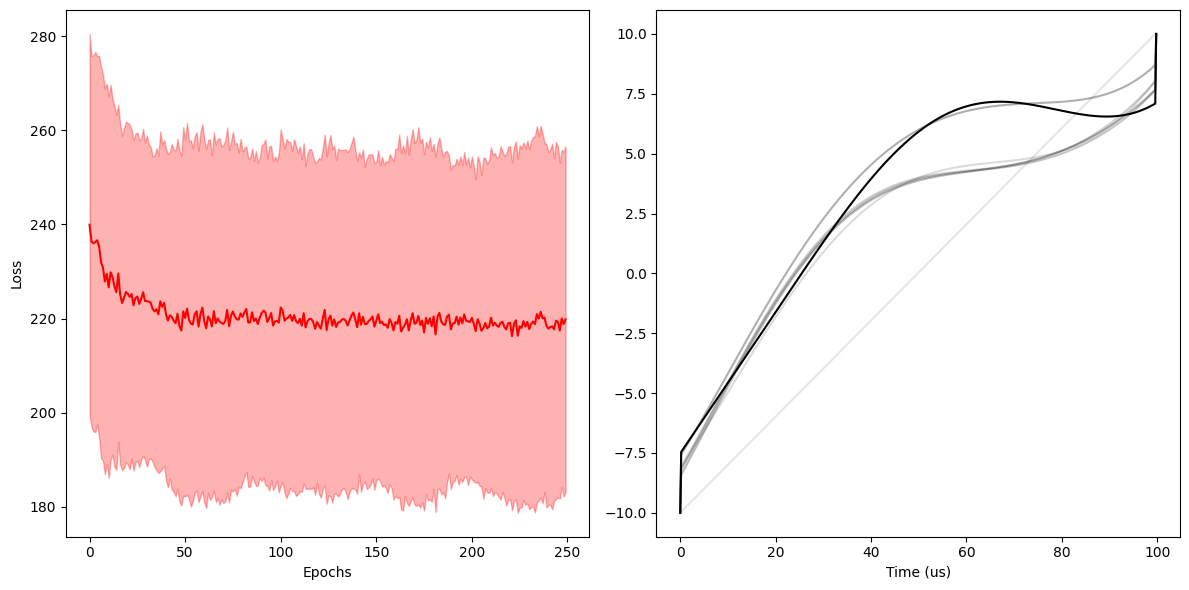}
    \caption{Loss convergence (left) and non-monotonic optimal protocol $\lambda_\theta(t)$ derived using AD to minimize the work done according to the potential~\ref{eq:linear_biasing}, following reference~\cite{ZhongWorkDissipation2022}.}
    \label{fig:reproducing-zhong}
\end{figure}

As shown in Figure~\ref{fig:reproducing-zhong}, AD can indeed reproduce the nonmonotonic behavior of the optimal driving schedule for limited control on a linear biasing force as seen in~\cite{ZhongWorkDissipation2022}.

\label{appendix:reproduce_protocols}

\section{Alternative objective function}
\label{appendix:other_loss}
\begin{figure}[]
    \centering
    \vspace{1em}
    \includegraphics[width=0.7\linewidth]{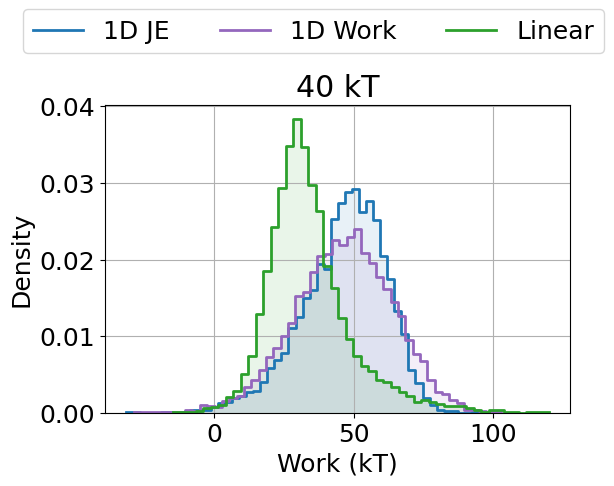}
    \caption{Non-Gaussian work distributions for 300\si{\micro\second} simulations with a 40$\kT$ landscape barrier. The 1D control work-minimized protocol (purple) yields a work distribution with a (dimensionless) skewness of $1.17$, while the 1D control protocol that minimizes Jarzynski error as expressed in equation~\ref{eq:je-loss} (blue) has a work distribution with skewness $-0.61$. The naive linear protocol work distribution (green) has skewness of $-0.20$.}
    \label{fig:je-error}
\end{figure}

Geiger and Dellago~\cite{geiger_optimum_2010} calculate protocols that minimize the error in free energy estimates:
\begin{equation}\label{eq:je-loss}
    \mathcal{L}(\lambda) = \langle \exp(\beta W_R)\rangle_\lambda
\end{equation} 
and find that this loss function yields improved free energy estimates compared to minimal-dissipation protocols. 

If the work distribution $p(W)$ is Gaussian, all cumulants of order 3 or higher vanish and the variance of the Jarzynski estimator, which is assumed to dominate the mean squared error in the free energy estimate~\cite{gore_bias_2003}, can be approximated by  $\mathrm{Var}(\exp(-\beta W)) \sim \exp(-2\beta W_{\mathrm{dis},F})$~\cite{geiger_optimum_2010}, which further simplifies to Equation~\ref{eq:je-loss} upon use of the Crooks fluctuation theorem~\cite{Lechner_2007}. 

\begin{figure*}[t]
    \centering
    \includegraphics[width=0.31\textwidth]{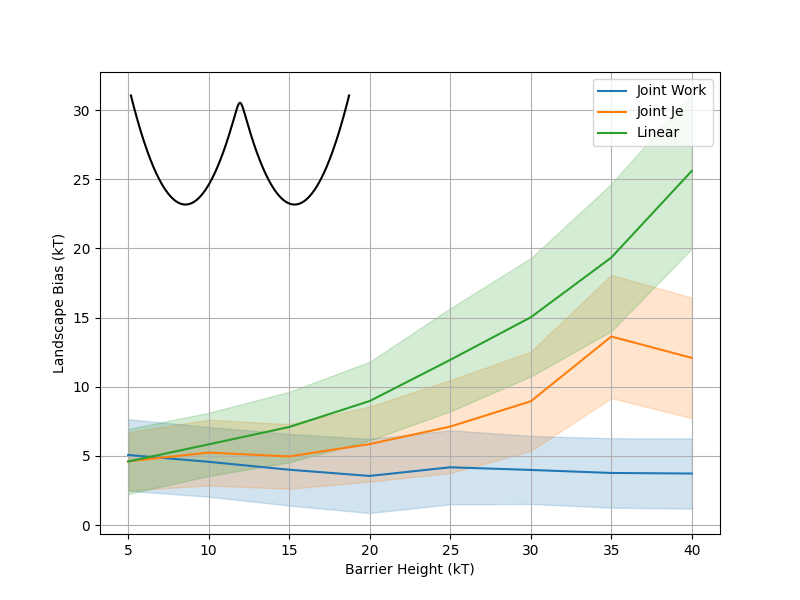}    
    \includegraphics[width=0.31\textwidth]{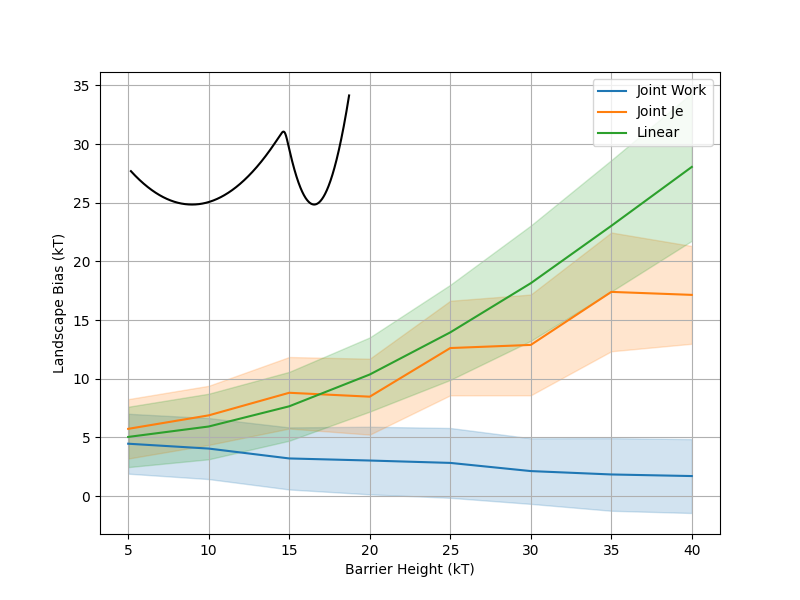}
    \includegraphics[width=0.31\textwidth]{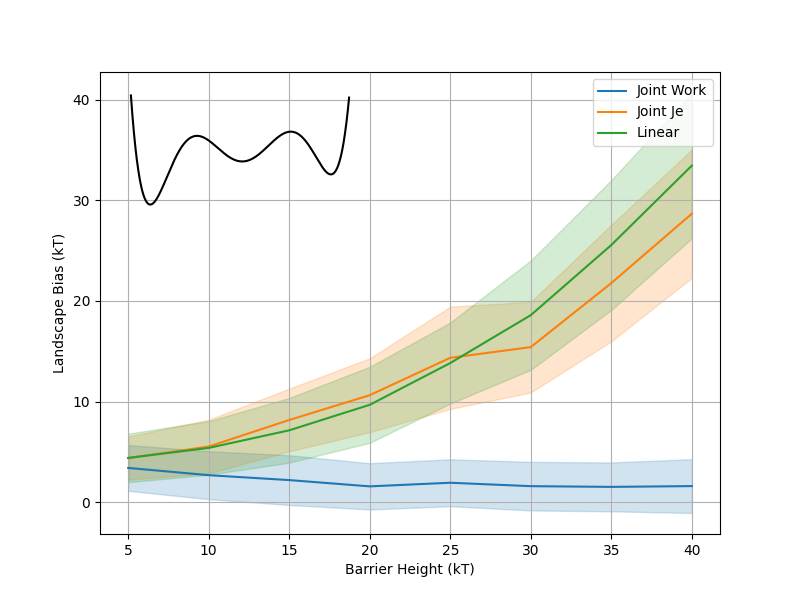}
    \caption{Bias (error) in the iteratively reconstructed landscapes as a function of barrier height. Simulation times are 0.0003, 0.0001, and 0.00003 from left to right, respectively. \emph{Joint} protocols utilize 2D control. \emph{je} (orange) corresponds to minimizing the loss from Equation~\ref{eq:je-loss}, and \emph{work} (blue) corresponds to minimizing work directly. The results of using a linear protocol are in green. The protocols that minimize Jarzynski error as expressed in Equation~\ref{eq:je-loss} clearly do not improve the quality of reconstructions, likely due to non-Gaussian work distributions in this regime and corresponding inapplicability of the derivation of Equation~\ref{eq:je-loss}~\cite{geiger_optimum_2010}.}
    \label{fig:error_barrier_height}
\end{figure*}

When driving far from equilibrium, our work distributions are non-Gaussian (see Figure~\ref{fig:je-error}). This invalidates the assumptions used to derive the loss in equation \ref{eq:je-loss}. As seen in Figure~\ref{fig:error_barrier_height}, we find that far from equilibrium, protocols optimized with Equation~\ref{eq:je-loss} perform poorly, even with 2D control.

\end{document}